\documentclass{aa}
\usepackage{graphicx}
\usepackage{natbib}
\usepackage{epstopdf}
\usepackage{supertabular}
\bibpunct{(}{)}{;}{a}{}{,}   % A&A style
\newcommand{\lesssim}{\mathrel{\hbox{\rlap{\hbox{\lower4pt\hbox{$\sim$}}}\hbox{$<$}}}}
\newcommand{\grsim}{\mathrel{\hbox{\rlap{\hbox{\lower4pt\hbox{$\sim$}}}\hbox{$>$}}}}
\newcommand{\gammamax}{\gamma_{\rm max}}
\newcommand{\gammamin}{\gamma_{\rm min}}
\newcommand{\gcool}{\gamma_{\rm cool}}
\newcommand{\taus}{\tau_{\rm s}}
\newcommand{\gammap}{\gamma_{\rm p}}
\newcommand{\eqb}{\begin{eqnarray}}
\newcommand{\eqe}{\end{eqnarray}}
\newcommand{\tesc}{t_{\rm esc}}
\newcommand{\tcool}{t_{\rm cool}}
\newcommand{\diff}{{\rm d}}
\newcommand{\sigmaT}{\sigma_{\rm T}}

\newcommand{\nelec}{N_{\rm e}}
\newcommand{\diffn}{n_{\rm e}}
\newcommand{\tauT}{\tau_{\rm T}}

\newcommand{\nuL}{\nu_{\rm L}}
\newcommand{\doppler}{{\cal D}}

\newcommand{\nuabs}{\nu_{\rm abs}}
\newcommand{\nuGHz}{\nu_{\rm GHz}}
\newcommand{\nusynch}{\nu_{\rm s}}
\newcommand{\numax}{\nu_{\rm max}}
\newcommand{\nuobs}{\nu_{\rm obs}}
\newcommand{\nup}{\nu_{\rm p}}
\newcommand{\nucool}{\nu_{\rm cool}}

\begin{document}
\title{A synchrotron self-Compton model with low energy electron cut-off for the blazar S5 0716+714}
\author{Olivia Tsang
\and J. G. Kirk
}
\institute{Max-Planck-Institut-f\"ur Kernphysik, Saupfercheckweg 1, 
D-69117 Heidelberg, Germany
}
\offprints{O. Tsang, \email{olivia.tsang@mpi-hd.mpg.de}}
\date{Received \dots / Accepted \dots }
\abstract 
{%context
  In a self-absorbed synchrotron source with power-law electrons,
  rapid inverse Compton cooling sets in when the brightness
  temperature of the source reaches $T_{\rm
    B}\sim10^{12}\,$K. However, brightness temperatures inferred from
  observations of intra-day variable sources (IDV) are well above the
  "Compton catastrophe" limit. This can be understood if the
  underlying electron distribution cuts off at low energy.}  
  {%aims
  We examine the compatibility of the synchrotron and inverse Compton
  emission of an electron distribution with low-energy cut-off with
  that of IDV sources, using the observed spectral energy distribution
  of S5~0716+714 as an example.}  
  {%methods
  We compute the synchrotron self-Compton (SSC) spectrum of
  monoenergetic electrons and compare it to the observed spectral
  energy distribution (SED) of S5~0716+714. The hard radio
  spectrum is well-fitted by this model, and the optical data can be
  accommodated by a power-law extension to the electron spectrum. We
  therefore examine the scenario of an injection of electrons, which is
  a double power law in energy, with a hard low-energy component that
  does not contribute to the synchrotron
  opacity.}
  {%results 
  We show that the double power-law injection model is in good
  agreement with the observed SED of S5~0716+714. For intrinsic
  variability, we find that a Doppler factor of $\doppler\geq30$ can
  explain the observed SED provided that low-frequency ($<32\,$GHz)
  emission originates from a larger region than the higher-frequency
  emission. To fit the entire spectrum, $\doppler\ge65$ is needed. We find
  the constraint imposed by induced Compton scattering at high $T_{\rm
    B}$ is insignificant in our model.  }
{%conclusions
  We confirm that electron distribution with a low-energy cut-off can
  explain the high brightness temperature in compact radio sources. We
  show that synchrotron spectrum from such distributions naturally
  accounts for the observed hard radio continuum with a softer optical
  component, without the need for an inhomogeneous source. The required low energy electron distribution is compatible with a relativistic Maxwellian.}
\titlerunning{SSC model for S5 0716+714}
\authorrunning{O. Tsang and J. G. Kirk}
\maketitle

\section{Introduction}
\label{intro}

Observations of many extra-galactic radio sources have found rapid
flux variations at radio frequency
\citep[e.g.][]{kedziorachudczeretal01}, some of which fluctuate over a
time scale of a day or less. They are referred to as intra-day variable
sources (IDV). The variability time scale is often used to constrain
the size of the source based on causality arguments. Using this
constraint, one can derive a variability brightness temperature
\citep{wagnerwitzel95} 
\eqb T_{\rm
  var}=4.5\times10^{10}F_{\nu}\left({\lambda d_{\rm L}\over t_{\rm
      obs} (1+z)}\right)^2\, {\rm K} 
\eqe 
where the flux density $F_\nu$, wavelength $\lambda$, luminosity distant $d_{\rm L}$, and
observed variability time scale $t_{\rm obs}$ are measured in Jy, cm,
Mpc, and days, respectively.

The high radio flux frequently measured in IDV sources implies
an extremely high brightness temperature, often many orders of magnitude
above $10^{12}\,$K. \citet{kellermannpaulinytoth69} have shown that,
assuming the electron distribution follows a single power law, the
luminosity of the inverse Compton scattered photons exceeds that of
the synchrotron photons when the brightness temperature of the source
reaches $\sim10^{12}\,$K. Above this threshold, rapid cooling of the
relativistic electrons due to inverse Compton scattering --- the  \lq\lq Compton
catastrophe\rq\rq\ --- forbids a further increase in the brightness temperature
\citep[see e.g.][for a recent review of the brightness temperature
problem]{kellermann02}. The limiting value is even lower, $T_{\rm
  B}<10^{11}\,$K, if the magnetic field and particle energy density of
the source is driven towards equipartition \citep{readhead94}. The
observed variability in some sources can be interpreted as the result of
extrinsic effects, which, at first sight, relaxes the size constraint. 
For example, the flux variations of PKS~1519$-$273
and PKS~0405$-$385 are convincingly identified as interstellar
scintillation. Nevertheless, all realistic models of the scintillation
mechanism impose a new constraint on the size and require a brightness 
temperature of $T_{\rm B}>10^{13}\,$K in some 
cases \citep{macquartetal00,rickettetal02},
far exceeding the limit imposed by the Compton catastrophe.

A prevalent feature associated with IDV sources is a flat or inverted
spectrum ($\alpha\leq0$, with flux $F_\nu\propto\nu^{-\alpha}$) at
radio-millimeter wavelengths
\citep[e.g.,][]{gearetal94,kedziorachudczeretal01}. Optically thick
synchrotron emission from power-law electrons rises as $\nu^{5/2}$,
too fast to account for the observed spectra. Optically thin
synchrotron emission in the scope of the conventional interpretation
of the synchrotron theory has a flux $F_\nu\propto\nu^{-(s-1)/2}$,
where $s$ is the power-law index of the electrons
($\diff\nelec/\diff\gamma\propto\gamma^{-s}$). If
$\alpha=(s-1)/2\leq0$, the number density of electrons diverges
towards high $\gamma$. Imposing a high-energy cut-off in the electron
spectrum avoids the divergence and may account for the commonly
observed spectral steepening at optical frequencies, but
\citet{marscher77} showed that electron spectra with $s\leq1$ would
result in a high flux between infrared and optical frequencies that is
not supported by observations. The most common interpretation of the
flat or inverted spectra is, therefore, a superposition of many
synchrotron spectra within an inhomogeneous source
\citep[e.g.][]{debruyn76,marscher77,blandfordkoenigl79}.

In \cite{kirktsang06}, we discussed a synchrotron self-Compton model in
which the electron distribution is monoenergetic. The lack of low-energy electrons enables more GHz photons to emerge from the source,
allowing a higher brightness temperature to be observed without
initiating catastrophic cooling. We found that a temperature of up to
$T_{\rm B}\sim10^{14}\,$K at GHz frequencies is possible with only a
moderate Doppler boosting factor of $\sim10$. In \citet{tsangkirk07},
we discussed the parameters of the monoenergetic model and showed
that the assumption of equipartition of energy in the source does not prevent the
Compton catastrophe. We also showed that an injection of highly
relativistic electrons or strong acceleration in the source cannot
produce temperatures much higher than our limit due to copious
electron-positron pair production.

In this paper, we examine the spectral properties of synchrotron
emission from monoenergetic electrons and from an electron
distribution that is a double power law in energy, by comparing the
model spectra with the observations of S5~0716+714, a BL~Lac object that is one of the brightest known IDV sources, as well as a gamma-ray blazar
\citep{hartmanetal99}.   
In doing so, we assume that the dominant targets for inverse Compton scattering
are produced within the source (SSC model).  
The emission from gamma-ray blazars can also be interpreted in the context of models in which 
the target photons are created externally (EC model), for example in the broad line region, 
the accretion disk, or a molecular torus \citep{sokolovmarscher05}. 
However, in many sources there is no observational evidence of a significant external
photon source. This is the case for S5~0716+714, where, 
despite much effort over the past three decades, no emission lines have been
  detected \citep[e.g.,][]{bychkova06}. 
 Furthermore, XMM-Newton
  observations of S5~0716+714 in 2004 analysed by \citet{ferreroetal06}
  and \citet{foschinietal06} show two spectral components in
  the $0.5-10\,$keV band, whose variability properties appear to favour the SSC interpretation.
The recent extensive simultaneous observations of this object from radio to
  optical frequencies by \citet{ostoreroetal06}, together with INTEGRAL
  pointings at GeV $\gamma$-ray energies during the same period, 
provide the best test for our model.

In the following, we present the computation of the
stationary electron distribution and the resulting synchrotron and
inverse Compton spectra. The model spectra computed using the
monoenergetic electron approximation, as described in
\citet{tsangkirk07}, are presented first. Although adequate for the radio emission, 
the monoenergetic model cannot reproduce the entire spectrum
of S5~0716+714. We therefore investigate an electron distribution that is
a double power law in energy --- a hard low-energy part that
softens to a high-energy tail above a characteristic energy. In this way, the
inverted optically thin radio emission is retained and complemented by
nonthermal synchrotron emission from the high energy tail. In section~\ref{parameter}, we briefly describe these injection models. The resulting stationary
electron distribution is calculated in section~\ref{stationarysoln}
and used for the computation of the synchrotron and inverse
Compton spectra. In section~\ref{sed}, we compare the predictions
  of these models with the observed spectral energy distribution (SED)
  of the source to S5~0716+714. 
Our findings and some limitations of our approach are
  discussed in section~\ref{discussion} and our conclusions 
presented in section~\ref{conclusion}.

\section{The model}
\label{parameter}

The homogeneous monoenergetic model discussed previously
\citep{kirktsang06, tsangkirk07} can be completely characterised by
the Doppler boosting factor
$\doppler=1/[\Gamma(1-\beta\cos\vartheta)]$ ($c\beta$ is the
source speed with respect to the rest frame of the host galaxy,
$\vartheta$ the angle between the velocity and the line of sight,
and $\Gamma=(1-\beta^2)^{-1/2}$),
the redshift of the host galaxy $z$, and four source parameters, the
electron number density $\nelec$, the magnetic field strength $B$, the
linear size of the source $R$, and the electron Lorentz factor measured in 
the rest frame of the source
$\gamma$. For the purpose of comparison with observations, these can
be transformed into a different set of parameters. Details of the
transformation can be found in \citet{kirktsang06}, in which $\nelec$,
$B$, and $\gamma$ are replaced by the characteristic frequency of
synchrotron emission, $\nusynch=\gamma^2\nu_0$, where $\nu_0=3eB/(4\pi
mc)$, the Comptonisation parameter $\xi$, which is the ratio of the luminosity of each successive generation of inverse Compton scattered photons to the luminosity of the previous
generation: $\xi=4\gamma^2\tauT/3$, (where $\tauT=\nelec R\sigmaT$ is
the Thomson optical depth), and the optical depth $\taus$ to
synchrotron self-absorption at the observing frequency. The size of
the source, $R$, can be constrained, for example, by applying
causality arguments to the variation time, $\Delta t$, of the source:
$R<c\Delta t\doppler/(1+z)$.

\begin{figure}
\includegraphics[width=8.5 cm]{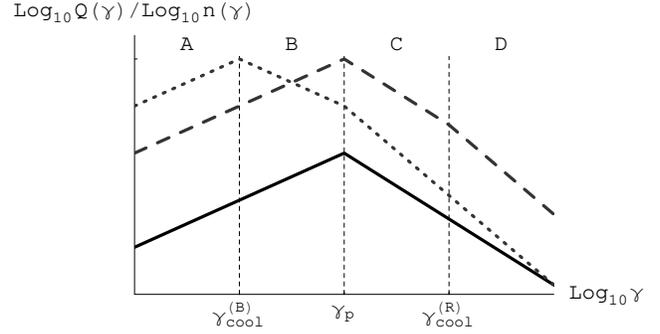}
\caption{\label{figne} Schematic representation of the electron injection spectrum and the stationary differential number density as a function of $\gamma$. The heights of the spectra have been adjusted for easy comparison and are not to scale. The solid line shows the double power-law injection spectrum with power-law index $s_1$ for $\gamma<\gammap$, and $s_2$ for $\gamma>\gammap$. The dashed line shows the case where $\gcool=\gcool^{\rm(R)}>\gammap$. The differential electron number density $\diffn\propto\gamma^{-s_1}$ in regions A and B, $\diffn\propto\gamma^{-s_2}$ in region C, and $\diffn\propto\gamma^{-(s_2+1)}$ in region D. The dotted line shows the case where $\gcool=\gcool^{\rm(B)}<\gamma_0$, with $\diffn\propto\gamma^{-s_1}$ in region A, $\diffn\propto\gamma^{-2}$ in region B, and $\diffn\propto\gamma^{-(s_2+1)}$ in regions C and D.}
\end{figure}

We present in Section~\ref{sed} the model spectra from monoenergetic
electrons that show good agreement with the observations of
S5~0716+714 at radio frequencies. The optical data can be fitted by
this model if a high-energy power-law \lq\lq tail\rq\rq\ is added.
To do this, we consider an injection spectrum of the form
$Q(\gamma)\propto(\gamma/\gammap)^{-s}$ for
$\gammamin<\gamma<\gammamax$, where the power-law index $s$ equals
$s_1$ for $\gamma<\gammap$, and $s_2$ for $\gamma>\gammap$ (Fig.~\ref{figne}). The electron number density at a
given time is proportional to $\gammamax^{1-s}$ for $s<1$,
$\propto\gammamin^{1-s}$ for $s>1$, and $\propto\ln\gammamax$ for
$s=1$. In the high-energy branch of the injection spectrum, for
$\gamma>\gammap$, we require that $s_2>1$, so that electron number density
congregates towards $\gammap$. In the low-energy branch,
$\gamma<\gammap$, the electrons congregate at $\gammap$ if
$s_1<1$. But we also require that the opacity to synchrotron
self-absorption is dominated by electrons with $\gamma=\gammap$, which
is achieved by demanding $s_1<1/3$. Under the conditions $s_1<1/3$ and
$s_2>1$, the low-frequency synchrotron spectrum is well-approximated
by that of monoenergetic electrons with Lorentz factor $\gammap$.

The electron injection spectrum cuts off at $\gammamin$ towards low
energy and at $\gammamax$ towards high energy. The exact value of
$\gammamin$ is unimportant, since, as explained above,
synchrotron emission and opacity are dominated by electrons with
$\gamma=\gammap$ in the low-energy part of the injection
spectrum, where $\gammamax$ determines the high frequency cut-off in the
synchrotron spectrum, at $\numax=\gammamax^2\nu_0$, and the highest
photon energy achievable through inverse Compton scattering in the
Klein-Nishina limit, which equals $\gammamax mc^2$.

To summarise, the injection spectrum has the form 
\eqb Q(\gamma)= Q_0
\left\{\begin{array}{cc}{\left({\gamma/\gammap}\right)^{-s_1}},\, 
& \gammamin\leq\gamma<\gammap \\
    \\
    {\left({\gamma/\gammap}\right)^{-s_2}},\, &
    \gammap\leq\gamma<\gammamax\end{array}\right.
\label{elecinj}
\eqe
where $Q_0$ is the electron injection rate per unit volume per unit
$\gamma$ at $\gamma=\gammap$.

\section{Stationary solution}
\label{stationarysoln}

The shape of the synchrotron spectrum is determined by the
stationary electron-energy distribution. Electrons injected
into the source according to Eq.~(\ref{elecinj}) are 
subject to radiative cooling while in the source and evacuate
this zone on a time-scale close to the light crossing time,
$t_{\rm esc}\sim R/c$. The evolution of the electron spectrum is
governed by the kinetic equation \citep{kardashev62}: 
\eqb
{\partial\diffn\over\partial t}&=&
Q_0\left({\gamma\over\gammap}\right)^{-s}
-{\partial\over\partial\gamma}\left(\diffn\dot{\gamma}_{\rm total}\right)
-{\diffn\over\tesc}
\label{kineq}
\eqe 
where, for simplicity, we denote the differential
electron number density $(\diff\nelec/\diff\gamma)$ by $\diffn$. 
The second term on the right hand side of Eq.~(\ref{kineq}) is the rate of change of 
the electron Lorentz factor due to radiative losses. This term is the sum of 
the rates for synchrotron losses and for losses from inverse Compton scattering:
\eqb
\dot{\gamma}_{\rm total}&=&
\dot{\gamma}_{\rm s}+\dot{\gamma}_{\rm IC}
\eqe 
where
\eqb
\dot{\gamma}_{\rm s}&=&{4\sigma_T U_{\rm B}\over 3mc}\gamma^2.
\eqe 
The third term is the rate at which electrons escape from the source.

In the stationary state, Eq.~(\ref{kineq}) can be solved analytically:
\eqb 
\diffn(\gamma)={1\over f_{\rm
    I}(\gamma)}\int^{\gamma}{Q\left(\gamma'\!'\right)\over\dot{\gamma'\!'}_{\rm total}}
f_{\rm I}\left(\gamma'\!'\right)\diff\gamma'\!'
\label{gensoln}
\eqe
with the integrating factor
\eqb
f_{\rm I}(\gamma)=
\dot{\gamma}\,{\rm exp}\left[-\int^{\gamma}
\left(\dot{\gamma'}_{\rm total}\tesc\right)^{-1}\diff\gamma'\right].
\eqe

However, Eq.~(\ref{kineq}) is only a rough description of a source, for example, 
because of the crude treatment of particle escape involved in setting $\tesc=R/c$.
Therefore, rather than use Eq.~(\ref{gensoln}), we choose to use an approximate 
solution that more clearly demonstrates the effects
{\bf that} cooling and the evacuation of electrons from the emission
region have on the electron energy distribution. 

We first identify the Lorentz factor, $\gcool$, which determines the electron
energy at which radiative cooling dominates losses due to
particles escaping the emission region:
\eqb
\left.{\dot{\gamma}_{\rm total}\over\gamma}\right|_{\gamma=\gcool}&=&
{1\over\tesc}\, .
\label{gcooldef}
\eqe
In principle, $\gamma_{\rm cool}$ can be evaluated only if the entire electron distribution is
already known, since $\dot{\gamma}_{\rm IC}$ depends on the spectrum and intensity of 
emitted radiation. However, in practise, a simple iterative scheme enables it to 
be found rapidly in all the cases we have computed. 
Assuming it is known, solutions of Eq.~(\ref{kineq}) that are valid in the limits
$\gamma\ll\gamma_{\rm cool}$ and $\gamma\gg\gamma_{\rm cool}$ are easily found.
In the first case, cooling is unimportant, and it immediately follows that
\eqb
\diffn&=&\tesc Q_0\left(\gamma/\gammap\right)^{-s}
\qquad\textrm{for }\gamma\ll\gamma_{\rm cool}.
\label{nocooling}
\eqe
In the second, escape is unimportant, and
the appropriate solution is found by integrating the kinetic equation once:
\eqb
\diffn&=&\dot{\gamma}^{-1}\int_{\gamma}^{\infty} Q_0\left({\gamma'/\gammap}\right)^{-s}
\diff\gamma'
\qquad\textrm{for }\gamma\gg\gamma_{\rm cool}.
\label{noescape}
\eqe
These solutions intersect close to the point $\gamma=\gamma_{\rm cool}$. 
Our approximation consists in adopting the solution without cooling 
given in Eq.~(\ref{nocooling}) for all Lorentz factors below the intersection point and
the solution without escape given in Eq.~(\ref{noescape}) for all Lorentz factors above the 
intersection point. 

In addition, we assume and verify {\em a posteriori} (see Section~\ref{discussion}) 
that $\dot{\gamma}_{\rm IC}$ can be approximated by the expression for inverse Compton scattering of the synchrotron photons in the Thomson regime:
\eqb
\dot{\gamma}_{\rm IC}&=&
{4\sigma_T U_{\rm s}\over 3mc}\gamma^2
\eqe
where $U_{\rm s}$ is the energy density of synchrotron photons in the source. 
In this case, $\dot{\gamma}_{\rm total}=\gamma^2/\left(\gamma_{\rm cool}\tesc\right)$,
and our approximate solution is
\eqb 
\diffn&=&\left\{
\begin{array}{cc}
\tesc Q(\gamma) & \gamma<a\gcool \\ \\
\tesc\gcool\gamma^{-2} \int_\gamma^\infty \diff\gamma'Q(\gamma')
& a\gcool\leq\gamma
\end{array}
\right.
\label{elecspec}
\eqe
where $a$ ($\sim1$) is determined by requiring the solution (but not its first derivative) 
to be continuous. 

The Lorentz factors $\gcool$ and $\gammap$ give rise to breaks in $\diffn$, which
correspond to the breaks in the synchrotron spectrum at $\nup=\gammap^2 \nu_0$ and $\nucool=\gcool^2 \nu_0$. Notice that, if $s<1$ (as in the injection spectrum below $\gammap$), $\diffn$ is approximately proportional to $\dot{\gamma}^{-1}\propto\gamma^{-2}$, whereas if $s>1$ (as in the injection spectrum above $\gammap$), $\diffn$ is approximately $\propto\gamma^{-(s+1)}$.

Two types of stationary spectra result from Eq.~(\ref{elecspec}), depending on whether the peak of the injection spectrum, $\gammap$, is below or above $\gcool$. Figure~\ref{figne} shows the injection spectrum as a solid line, the stationary spectra where $\gammap>\gcool$ as a dotted line and where $\gammap<\gcool$ as a dashed line. When electrons are predominantly removed from a certain energy range by leaving the source ($\tesc<\tcool$), the spectrum retains its original shape, $\diffn\propto\gamma^{-s}$, since $\tesc$ is independent of particle energy. On the other hand, when synchrotron losses dominates, such that $\tesc>\tcool$, the stationary solution is $\diffn\propto\gamma^{-2}$ for $\gamma<\gammap$, and $\diffn\propto\gamma^{-(s+1)}$ for $\gamma>\gammap$.  For the computation of the low frequency synchrotron emission, the distribution can be approximated by a monoenergetic one at $\gcool$ in the first case and $\gammap$ in the second.

The iterative procedure used to find $\gcool$ is as follows: The loss rate is defined as
\eqb
\dot{\gamma}_{\rm total}&=&{4\sigma_TU_{\rm B}(1+\delta)\over 3mc}\gamma .
\eqe
Then, starting with $\delta=0$, $\gcool$ is evaluated from Eq.~(\ref{gcooldef}) and, using 
the electron distribution given by (\ref{elecspec}), $U_{\rm s}$ is evaluated as described 
in Sect. \ref{synch_and_ic_emission}. The value of $\delta$ is readjusted to
$\delta=U_{\rm s}/U_{\rm B}$ and the cycle repeated until successive values differ 
by less than 1\%.  In the examples discussed in this paper, convergence was achieved 
after two iterations. Because the change in $\gcool$ between iterations was 
only roughly a factor of 2, the final emission spectrum was close to that found 
using $\delta=0$.    

\subsection{Synchrotron and inverse Compton emission}
\label{synch_and_ic_emission}
The synchrotron specific intensity, following straightforwardly from
the radiative transport equation, is \eqb I_\nu^{\rm
  (S)}=S_\nu\left[1-{\rm exp}(-\taus)\right]
\label{syncintdef}
\eqe where the optical depth to synchrotron radiation is
$\taus=\alpha_\nu\cdot R$, and $\alpha_\nu$ is the absorption
coefficient \citep[e.g.,][Chapter 18]{longair92} 
\eqb
\alpha_\nu=-&&\!\!\!\!{3\sqrt{3}\over16}{\sigmaT\over\alpha_{\rm
    f}}{mc^2\over h\nu}{\nuL\sin\phi\over\nu}
\nonumber \\
& &\,\,\times\int_{\gamma_{\rm min}}^{\gamma_{\rm
    max}}\gamma^2F(x){\diff\over\diff\gamma}\!
\left(\!{\diffn(\gamma)\over\gamma^2}\!\right)\diff\gamma
\eqe 
where $\alpha_{\rm f}$ is the fine structure constant, $\phi$ the angle between the magnetic field and the direction of the emitted radiation, $x=\nu/(\gamma^2\nu_0)$, $F(x)=x\int_x^{\infty}K_{5/3}(t)\diff t$, and $K_{5/3}$ is the modified Bessel function of order $5/3$. The source function $S_\nu$ is 
\eqb
S_\nu=-{2m\nu^2}{\int_{\gamma_{\rm min}}^{\gamma_{\rm max}}F(x)
  \diffn(\gamma) \diff \gamma\over\int_{\gamma_{\rm min}}^{\gamma_{\rm
      max}}\gamma^2 F(x)
  {\diff\over\diff\gamma}\left({\diffn(\gamma)\over\gamma^2}\right)
  \diff \gamma} .
\eqe 
In the monoenergetic approximation, the source function simplifies to 
\eqb 
S_\nu=m\nu_0^2}{\gamma^5F(x)\over K_{5/3}(x) .
\eqe
Equation~(\ref{syncintdef}) is integrated over frequency and angle to give 
the energy density of synchrotron photons in the source
\eqb
U_{\rm s}&=&{4\pi\over3c}\zeta\int_0^{\infty} I_\nu\diff\nu
\eqe
where $\zeta$ is a geometrical factor that is shown in \citet{tsangkirk07} to be $\zeta=2/3$.

The synchrotron photons are repeatedly scattered by the energetic electrons to higher energies. Denoting by $i$ the number of times a photon is scattered, the rate of scattering the $(i-1)^{\rm th}$ generation of photons into the frequency interval $\diff\nu_i$ by a single electron \citep[see e.g.,][Eq.~(4)]{georganopoulosetal01} is
\eqb
\left({\diff n_{\rm ph}\over\diff t\diff\nu_i}\right)_{\rm sp}={3\sigmaT c\over4\nu_{i-1}\gamma^2}f(y) N_{\nu_{i-1}}
\label{defscatrate}
\eqe
where
\eqb
N_{\nu_{i-1}}={4\pi\over c}{\zeta I_{\nu_{i-1}}\over h\nu_{i-1}}
\label{deftarph}
\eqe 
is the number density of the target photons, and $I_{\nu_{i-1}}$ the specific intensity of the $(i-1)^{\rm th}$ generation of photons. The first generation of scattered photons is produced directly from the synchrotron photons: $i=1$, $I_{\nu_0}=I_\nu^{\rm (S)}$. \citet[][Chapter 7]{rybickilightman79} assumed that scattering in the Thomson regime is isotropic in the rest frame of the electron, and obtained $f(y)\approx f_{\rm iso}(y)=2(1-y)/3$. Here, we include the Klein-Nishina effects \citep[e.g.,][]{georganopoulosetal01}, in which case 
\eqb
f(y)&=&\left[2\,y\,{\rm ln}\,y+y+1-2y^2+{(4\epsilon_{i-1}\gamma\,
    y)^2(1-y)\over2(1+4\epsilon_{i-1}\gamma\, y)}\right]
\nonumber \\
&&\times P(1/4\gamma^2,1,y)\, ,
\\
y&=&{\epsilon_i\over4\epsilon_{i-1}\gamma^2(1-\epsilon_i/\gamma)} 
\eqe
where $\epsilon_{i-1}$ and $\epsilon_i$ are the energy of the target photons and scattered photon, respectively, in units of $mc^2$, and $P(1/4\gamma^2,1,y)=1$ for $1/4\gamma^2\leq y \leq 1$, and zero
otherwise.

Assuming a spherical source, the rate of scattering photons with energy $h\nu_{i-1}$ to energy $h\nu_i$, in the observer's frame, from a homogeneous distribution of electrons with differential number
density $\diffn$ can be found by integrating over the electron energy distribution, 
\eqb 
\left({\diff n_{\rm ph}\over\diff t \diff
    \nu_i}\right)= {4\pi\over3}\left(R\over2\right)^3
\!\int_0^\infty\diff\gamma{\diffn} \left({\diff n_{\rm ph}\over\diff
    t\diff\nu_i}\right)_{\rm sp} .
\label{defscatrate2}
\eqe
Note that $R$ (the linear size of the source) is divided by 2 to obtain the source radius.

The specific intensity of the $i^{\rm th}$ generation photons is then the scattering rate of the electron distribution in Eq.~(\ref{defscatrate2}) integrated over all target photon frequency,
\eqb
I_{\nu_i}^{\rm (C)}&=&\left({\diff E\over\diff t\diff\nu_i\diff r^2\diff\Omega}\right)
\nonumber \\
&=&\int_0^{\infty}\diff\nu_{i-1}
\left({\diff n_{\rm ph}\over\diff t \diff \nu_i}\right){h\nu_i\over 4\pi (R/2)^2} ,
\eqe
and for a general electron distribution $\diffn$, it can be written as
\eqb
I_{\nu_i}^{\rm (C)}= {\pi\over3}R\sigmaT{\nu_i}
\!\int_0^\infty\!{\diff\gamma\over\gamma^2}{\diffn}\!
\int_0^\infty {\diff\nu_{i-1}\over\nu_{i-1}^2}I_{\nu_{i-1}}f(y) .
\label{powerlawdef}
\eqe
For a monoenergetic electron distribution, Eq.~(\ref{powerlawdef}) can be simplified to
\eqb
I_{\nu_i}^{\rm (C)}=
{\pi\over3}\tauT{\nu_i\over\gamma^2}\int_0^\infty {\diff\nu_{i-1}\over\nu_{i-1}^2}I_{\nu_{i-1}}f(y) .
\label{monodef}
\eqe
Equations~(\ref{monodef}) and (\ref{powerlawdef}) are integrated numerically for monoenergetic electrons and for an electron distribution given by Eq.~(\ref{elecspec}).

\section{The BL Lac object S5~0716+714}
\label{sed}

Observations of S5~0716+714 have shown that the source exhibits intra-day variability in the radio and optical bands \citep[e.g.][]{ghisellinietal97, raiterietal03}. Correlation between radio (at $5\,$GHz) and optical (at $650\,$nm) variability suggest that scintillation, a process that is not effective at high radio and
optical frequencies, does not play a large part in the observed variability \citep{quirrenbachetal91,wagner01}. More recent multi-frequency studies of S5~0716+714 \citep[e.g.][]{ostoreroetal06,agudoetal06} have obtained simultaneous measurements from radio to optical frequencies during the INTEGRAL pointing period, and the non-detection of the source by INTEGRAL has provided upper limits at X-ray frequencies. Flux variations were detected at 32 and $37\,$GHz over a period of $\Delta t=4.1\,$days (in November 2003) when the two-frequency measurements overlap. Since interstellar scintillation is ineffective at these frequencies, \citet{ostoreroetal06} conclude that the variability is intrinsic. Assuming $H_0=70\,{\rm km}\,{\rm sec}^{-1}\,{\rm Mpc}^{-1}$, with $\Omega_\lambda=0.7$, $\Omega_{\rm M}=0.3$, and $\Omega_{\rm k}=0$, and a redshift $z>0.3$ based on the non-detection of a host galaxy \citep[e.g.][]{wagneretal96}, they derive a variability brightness temperature of $T_{\rm var}>(2.1\pm0.1)\times10^{14}$ K.

\citet{bachetal05} analysed the data set of VLBI images of 11 jet
components of S5~0716+714 at 4.9 GHz, 8.4 GHz, 15.3 GHz, and 22.2 GHz,
observed between 1992 and 2001. Assuming that all the jet components
move with the same speed along the jet (i.e. all components have the
same bulk Lorentz factor), they propose that the observed wide range
(from 5.5$c$ to 16.1$c$) of apparent component speeds is due to
variations in the viewing angle and limit the Lorentz factor and the
viewing angle of the VLBI jet to $\Gamma>15$ and $\theta<2^{\circ}$,
respectively. Under these conditions, the range of Doppler factors would be
$\doppler\approx20-30$.

According to \citet{ostoreroetal06}, observations of S5~0716+714
between $5\,$GHz and $32\,$GHz can be fitted with spectral indices
$\alpha_{5-32}$ of $-0.3$ and $-0.5$ at two different epochs. They
suggest that the radio observations can be interpreted as optically
thick synchrotron emission from an inhomogeneous source, and the
spectral break at $\nu\approx10^{13}\,$Hz would correspond to the
self-absorption frequency $\nuabs$. In the near-infrared to optical
band, observations from 2001$-$2004, reported by
\citet{hagenthornetal06} suggest that the spectral energy
distribution between the frequencies $\nu_{\rm
  K}=1.38\times10^{14}\,$Hz and $\nu_{\rm B}=6.81\times10^{14}\,$Hz
can be fitted by the power law $F_\nu\propto\nu^{-1.12}$.

Here, we apply the two homogeneous models described in section~\ref{parameter}: the monoenergetic one, which successfully models the hard radio spectrum with relatively few free parameters, and 
the one with double power-law injection (and, consequently more free parameters), which also enables the high-energy emission to be modelled. We adopt a value of $z=0.3$ for the red-shift of S5~0716+714, and a linear size inferred by the variability time scale of $\Delta t=4.1\,$days, so that
$R=c\Delta t\doppler/(1+z)$. The values we find for the free parameters of the models 
and for several parameters derived from these are given in Table~\ref{tabvalues}. The spectra predicted by the two models are shown in Figs.~\ref{sedmono} and \ref{sedpowerlawall}, and are discussed
separately in the next two sections.

Identifying the creation of our homogeneous source with an event that leads to the ejection of an individual blob observed with VLBI, we estimate the minimum jet power implied by each set of parameters by multiplying the total energy content of the source by the average rate at which blobs are ejected from the core. In the co-moving frame of the source, for monoenergetic electrons, the total energy content is
\eqb 
E'_{\rm blob}=\left({B^2\over8\pi}+\nelec'\gamma
  mc^2\right){R'^3} 
\eqe 
and, for power-law electrons, 
\eqb 
E'_{\rm
  blob}=\left({B^2\over8\pi}+\int_{\gamma_{\rm min}}^{\gamma_{\rm
      max}}\diffn'(\gamma)\gamma mc^2\diff\gamma\right){R'^3} .
\eqe 
Transforming to the rest frame of the galaxy, 
$E_{\rm blob}=\Gamma E'_{\rm blob}$, so
that the jet power 
\eqb 
P_{\rm jet}&=&E_{\rm blob}\left({1\over\Delta
    t_{\rm ej}}\right) ,
\eqe 
where $\Delta t_{\rm ej}$ is the average time difference between the ejection of successive blobs. Linear fits of the change in the position of the 11 jet components \citep{bachetal05} suggest that a new component is ejected from the core every $0.1-1.8\,$years. The lower limits to the jet power of each model given in Table~\ref{tabvalues} were computed using $\Delta t_{\rm ej}=1.8\,$years.

\subsection{Monoenergetic electrons}

\begin{figure}
\includegraphics[width=8.5cm]{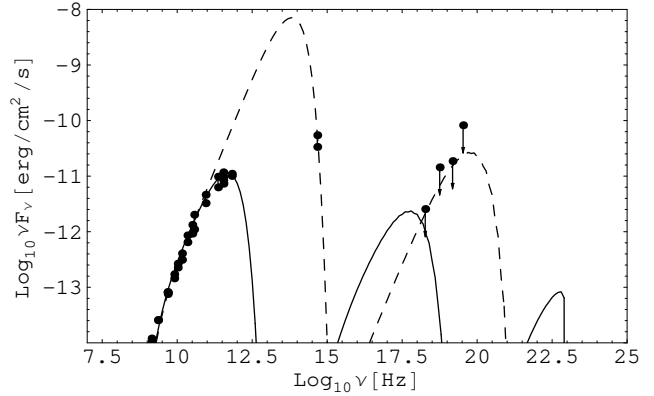}
\caption{\label{sedmono} Spectral energy distribution of S5~0716
  +714. Multi-frequency simultaneous data from \citet{ostoreroetal06}
  are shown as black symbols. Black dots show data points, variation
  ranges are shown by a vertical bar between two symbols, and downward
  arrows show upper limits. Values of the parameters are shown in
  Table~\ref{tabvalues}. The model spectra are computed from a
  distribution of monoenergetic electrons and are shown with solid
  and dashed lines. The dashed line shows the strong-magnetic-field model where
  the parameters are chosen such that it goes through the data points
  at optical frequency, whereas the solid line shows weak-magnetic-field model where the parameters are chosen to mimic the spectral
  turning at $10^{11.5}\,$Hz. The values of the parameters are shown
  in Table~\ref{tabvalues}.}
\end{figure}

In the monoenergetic model, the spectrum is specified by four parameters, $\nuabs$, $\nup$, $\doppler$, and $\xi$, as well as $z$ and $\Delta t$ (which are kept fixed for all models). The self-absorption frequency $\nuabs$ is determined by the first spectral break at $\sim4\,$GHz, and $\nup$ corresponds to
the spectral cut-off. In Fig.~\ref{sedmono}, we compare two models in which one has a cut-off at the spectral break at $\sim10^{11.5}\,$Hz, and the other cuts off just before reaching the optical point. The Doppler factor $\doppler$ affects the level of the observed flux both by determining the linear size of the source and determining the amount of boosting the flux receives. The Comptonisation parameter $\xi$ determines the ratio of the synchrotron flux to the inverse Compton flux, as well as the value of $\gammap$. Therefore, after specifying $\nuabs$ and $\nup$, $\xi$ must be adjusted to compensate for its effect on the level of the observed flux and to ensure the inverse Compton spectra do not exceed the INTEGRAL upper limits, while minimising $\doppler$.

Figure~\ref{sedmono} shows the measurements and variable ranges obtained from the simultaneous multi-frequency observation of S5~0716+714 from the study conducted by \cite{ostoreroetal06}. Also shown are the spectra predicted by the model assuming electrons are monoenergetic. The Doppler boosting factor is $\doppler=55$ in both models. The weak-magnetic-field model has a synchrotron self-absorption frequency of $\nuabs=3.9\,$GHz with a synchrotron spectrum that peaks at $\nup=300\,$GHz (the values of other parameters are shown in Table~\ref{tabvalues}). This model gives a brightness temperature of $T_{\rm B}=3.9\times10^{12}\,$K at $\nuobs=32\,$GHz ($T_{\rm B}=c^2F_\nu/(2k_{\rm B}\nu^2\theta_{\rm d}^2)$, where $k_{\rm B}$ is the Boltzmann constant and $\theta_{\rm d}$ the angular diameter of the source). The synchrotron spectrum shows good agreement with the data points at radio frequencies. The first-order inverse-Compton spectrum gives emission from optical to soft X-ray frequencies and the second-order spectrum gives gamma-ray emission up to $\sim40\,$MeV, while emission from higher orders scattering is negligible due to the Klein-Nishina effect.

The strong-magnetic-field model has $\nuabs=3.9\,$GHz, with its synchrotron peak at $\nup=55\times10^{12}\,$Hz (the values of other parameters can be found in Table~\ref{tabvalues}) and gives a brightness temperature of $T_{\rm B}=3.7\times10^{12}\,$K at the observing frequency $\nuobs=32\,$GHz. The synchrotron spectrum extends up to optical frequencies, and gives a reasonable fit at radio frequencies up to $\sim10^{11.5}\,$Hz. The first-order inverse-Compton spectrum gives X-ray emission, the second-order inverse-Compton spectrum is greatly reduced by the Klein-Nishina effects, and very little gamma-ray emission is produced.

The spectral break at $10^{11.5}\,$Hz is well-fitted by the weak-magnetic-field model. We are unable to obtain a set of parameters to allow the first inverse Compton spectrum to reproduce the optical data. Simple qualitative analysis shows that mimicking the optical data points with the first inverse Compton spectrum is inconsistent with observation. The level of flux that the first inverse-Compton spectrum would require in order to account for the optical data is much higher than the synchrotron flux (i.e. $\xi\gg1$), so a large $\gamma$ would therefore be required, resulting in the spectrum extending to frequencies far beyond the optical band. The first inverse-Compton spectrum would therefore exceed the INTEGRAL upper limits, and the very high X- and $\gamma$-ray flux would result in copious electron-positron pair production as the $\gamma$-ray photons interact with the synchrotron photons.

Alternatively, one can attempt to include the optical data in the synchrotron spectrum, as shown by the strong-magnetic-field model. The Lorentz factor of this model is higher than $\gcool$, which implies that the particles lose a significant portion of their energy by synchrotron radiation before they vacate the source, so the electron spectrum will evolve into one that is proportional to $\gamma^{-2}$. This set of parameters therefore violates the monoenergetic assumption. Furthermore, the predicted spectrum fails to account for the spectral break at $\nu\sim10^{11.5}\,$Hz, and the optical flux is very sensitively to the electron Lorentz factor. This model is, therefore, inconsistent. It is apparent that, in order to reproduce the observed optical emission, a power-law component in the electron spectrum at $\gamma>\gammap$ must be incorporated, which emits synchrotron radiation at a frequency above $\nup$.

\begin{table*}[t] 
\centering
\caption{{\bf Model parameters corresponding to} Figs.~\ref{sedmono} and \ref{sedpowerlawall}.}
\begin{tabular}{lllll}
\hline\hline \vspace*{-2mm} \\
& {\bf Mono (dashed)} & {\bf Mono (solid)} & {\bf Power-law (dashed)} & {\bf Power-law (solid)}
\vspace*{2mm} \\
 \hline \vspace*{-2mm} \\
{\bf Primary Parameters\hspace*{7mm} \vspace*{1mm}} & & & & \vspace*{3mm}\\
$z$ \vspace*{2mm} & 0.3 & 0.3 & 0.3 & 0.3 \\
$\Delta t$ (days) \vspace*{2mm} & 4.1 & 4.1 & 4.1 & 4.1\\
$\mathcal{D}$ \vspace*{2mm} & 55 & 55 & 30 & 65\\
$\nup$ (Hz) \vspace*{2mm} & $5.5\times10^{13}$ & $3.9\times10^{11}$ & $2.0\times10^{11}$ & $2.7\times10^{11}$ \\
$\nucool$ (Hz) & $1.74\times10^{12}$ & $1.17\times10^{19}$ & $2.07\times10^{16}$ & $2.14\times10^{19}$ \vspace*{2mm} \\
$\numax$ (Hz) & $\dots$ & $\dots$ & $1.0\times10^{18}$ & $1.5\times10^{15}$ \vspace*{2mm} \\
$\xi$ & $10^{-2.5}$ & $10^{-0.75}$ & $\dots$ & $\dots$ \vspace*{2mm} \\
\hline  \vspace*{-2mm} \\
{\bf Secondary Parameters} \vspace*{3mm} \\
$\xi$ & $\dots$ & $\dots$ & $0.92$ & $0.86$ \vspace*{2mm} \\
 $R$ (pc) & 0.15 & 0.15 & 0.08 & 0.17\vspace*{2mm}\\
 $\theta_{\rm d}$ ($\mu$as) & 32.5 & 32.5 & 17.7 & 38.4 \vspace*{2mm} \\
 $\gammap$ & 691 & 800 & 244 & 696 \vspace*{2mm} \\
 $\gcool$ & 123 & $4.39\times10^6$ & $7.85\times10^4$ & $6.19\times10^6$ \vspace*{2mm} \\
 $\gammamax$ & $\dots$ & $\dots$ & $5.45\times10^5$ & $5.19\times10^4$ \vspace*{2mm} \\
 $\nelec$ (cm$^{-3}$)& 0.02 & 0.70 & 3.15 & 0.32 \vspace*{2mm} \\
 $B$ (mG) & 648 & 3.43 & 34.7 & 2.65 \vspace*{2mm} \\
 $\delta$ & $\dots$ & $\dots$ & 0.97 & 0.93  \vspace*{2mm} \\
 $U_{\rm B}/U_{\rm par}$ & $1.8\times10^{3}$ & $1.0\times10^{-3}$ & 0.04 & $8.4\times10^{-4}$ \vspace*{2mm} \\
 $T_{\rm B}$ (K) & $3.7\times10^{12}$ & $3.9\times10^{12}$ & $1.4\times10^{13}$ & $2.5\times10^{12}$ \vspace*{2mm} \\
 $\tauT$ & $6\times10^{-9}$ & $2\times10^{-7}$ & $5\times10^{-7}$ & $1\times10^{-7}$ \vspace*{2mm} \\
 $P_{\rm jet}$ (ergs/s) & $1\times10^{45}$ & $3\times10^{43}$ & $5\times10^{42}$ & $4\times10^{43}$ \vspace*{2mm}\\
 \hline \\
\end{tabular}
\flushleft{Note: $\theta_{\rm d}$ is the angular diameter of the source, $U_{\rm par}$ the energy density of the particles, and $P_{\rm jet}$ is the jet power in the rest frame of the host galaxy, as predicted by each model. The compactness of all four models are negligibly small so not included in the discussion.}
\label{tabvalues}
\end{table*}

\subsection{Injection of electrons in a double power law}

\begin{figure*}
\centering
\includegraphics[width=12cm]{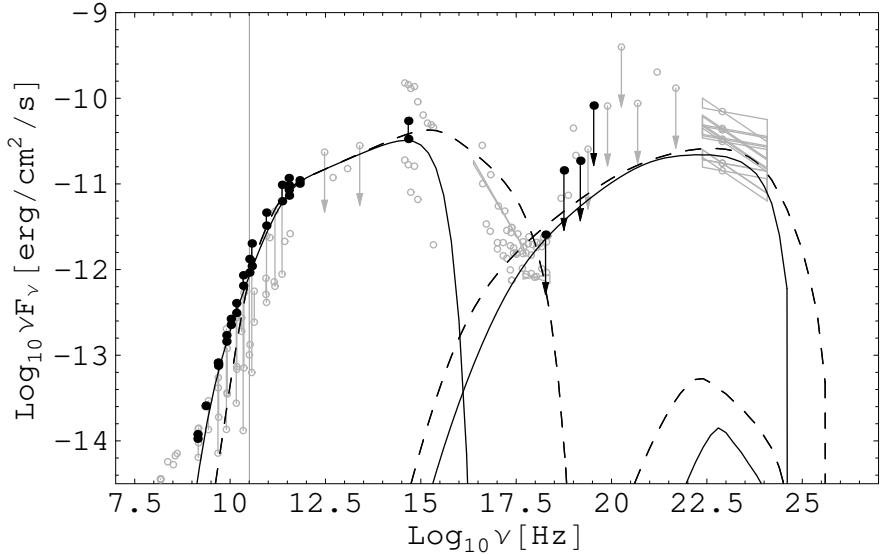}
\caption{The spectral energy distribution of S5~0716 +714, as represented in Fig.~\ref{sedmono}. The model spectra, shown as solid and dashed lines, are computed from a quasi-monoenergetic electron distribution in the form of Eq.~(\ref{elecspec}). The dashed line represents the low-Doppler-factor model where the Doppler boosting factor is minimised, whereas the solid line shows the high-Doppler-factor model where the values of the parameters are chosen to account for all radio and optical data points. The dashed gridline shows the position of $32\,$GHz. The values of the parameters are shown in Table~\ref{tabvalues}. Historical data, as compiled by \citet{ostoreroetal06} and shown as grey symbols,  %at the wavelengths of 1.38, 2.7, 3.9, 7.7, 13 and 31 cm are from RATAN-600; other 
at radio-to-optical frequencies are from \citet{kuehretal81}, \citet{waltmanetal81}, \citet{eckartetal82}, \citet{perley82}, \citet{perleyetal82}, \citet{lawrenceetal85}, \citet{saikiaetal87}, \citet{kuehrschmidt90}, \citet{moshiretal90}, \citet{halesetal91}, \citet{krichbaumetal93}, \citet{gearetal94}, \citet{halesetal95}, \citet{douglasetal96}, \citet{rengelinketal97}, \citet{zhangetal97}, \citet{rileyetal99}, \citet{cohenetal02}, \citet{raiterietal03}; UV data from \citet{piantreves93}, \citet{ghisellinietal97}; X-ray data from \citet{biermannetal92}, \citet{comastrietal97}, \citet{kuboetal98}, \citet{giommietal99}, \citet{tagliaferrietal03}, \citet{pianetal05}; and $\gamma$-ray data from \citet{mcnaronbrownetal95}, \citet{hartmanetal99}, \citet{collmar06}.}
\label{sedpowerlawall}
\end{figure*}

The Comptonisation parameter $\xi$ in this model must account for the $\gamma$ dependence of the 
electron density. It is therefore redefined as
\eqb
\xi={4\over3}R\sigmaT\int_0^\infty\gamma^2\diffn\diff\gamma ,
\label{defxi}
\eqe 
which is consistent with the definition used in the monoenergetic model.
However, for the purpose of fitting the flux in the INTEGRAL frequency 
range, $\sim10^{18}-10^{20}\,$Hz, we introduce a parameter $r_{\rm p}$, that determines the ratio of
flux between $\nup$ and $\gammap^2\nup$. Inspection of the solid line
in Fig.~\ref{sedmono} shows that synchrotron photons at the
spectral break $\nup\approx300\,$GHz are scattered to
$\gammap^2\nup\approx10^{17.5}\,$Hz. Therefore, $r_{\rm p}$ is a
parameter that allows us to specify the spectral break $\nup$ and then
to adjust the flux in the INTEGRAL frequency range. The normalisation
constant in the double power-law injection spectrum, $Q_0$ is
eliminated in favour of the parameter $r_{\rm p}$ 
\eqb 
r_{\rm
  p}={4\over3}\gammap^2R\sigmaT\int_0^\infty\diffn(\gamma)\diff\gamma ,
\label{defrp}
\eqe 
with $\diffn(\gamma)$ given by Eq.~(\ref{elecspec}). In the monoenergetic approximation, $r_{\rm p}$ is equivalent to $\xi$. Therefore, using the monoenergetic model and specifying $\nuabs$ and $\nup$, the radio and the X-ray flux can be adjusted with $\xi\equiv r_{\rm p}$ and $\doppler$, and $\nucool$
can be calculated.

In the low-Doppler-factor model (Figs.~\ref{sedpowerlawall} and \ref{sedpowerlawopt}), we attempt to minimise the Doppler factor of the source. According to \citet{wagneretal96} and \cite{ostoreroetal06},
the variability displayed by S5~0716+714 is intrinsic, and the variation time $\Delta t=4.1\,$days was measured at $32\,$GHz and $37\,$GHz. Therefore, we require that the model spectrum must agree with the data at these two frequencies. We are unaware of correlated simultaneous variability measurement at lower frequencies during this campaign and, therefore, allow the model spectrum to deviate from the data at frequencies below $32\,$GHz. At the expense of having a lower flux than what is observed at frequencies below $32\,$GHz, we find that the minimum Doppler factor required is reduced
to $\doppler=30$.

The power-law indices of the injection spectrum used to generate the spectrum of the low-Doppler-factor model are $s_1=-2$ for the low-energy part, such that electrons with $\gamma<\gammap$ do not contribute significantly to the synchrotron opacity, and $s_2=2.60$, chosen for constructing the spectral shape in the infrared to optical band. The rest of the parameters are varied while keeping the Doppler boosting factor fixed. To find the limiting case, we have chosen the self absorption frequency to be $\nuabs=32\,$GHz and find that the minimum Doppler factor that can generate a high enough level of flux at $32\,$GHz and beyond is $\doppler=30$. The values of the other parameters can be found in Table~\ref{tabvalues}. At the observing frequency of $32$ GHz, the brightness temperature in the frame of the observer is $T_{\rm B}=1.4\times10^{13}\,$K. The frequency at which the synchrotron
spectrum cuts off does not affect the spectral shape at low frequencies. However, $\numax=\doppler/(1+z)\times\nu_0\gammamax^2$ is constrained by the optical data, which imposes a lower limit on
$\numax$, and the INTEGRAL upper limits, which impose an upper limit on $\numax$. The maximum value is shown {\bf in this model}, where $\numax=10^{18}\,$Hz. This equates to $\gammamax=5.45\times10^5$ with $\doppler=30$ and $z=0.3$.

The low-Doppler-factor model shows that it is possible to interpret the observed variability at $32\,$GHz and $37\,$GHz as coming from one of the jet components with the kinematics
described by \citet{bachetal05}. That would require the lower-frequency emission to originates from a larger region than what is inferred from the observed variability at $\nu\geq32\,$GHz.

In the high-Doppler-factor model (Figs.~\ref{sedpowerlawall} and \ref{sedpowerlawopt}) we assume the emission at all frequencies --- including the low frequency ($<32\,$GHz) radio ---  originates from a single homogeneous source. This is suggested by the correlation between the variability at $5\,$GHz and $650\,$nm observed in February 1990 \citep{quirrenbachetal91,wagneretal96}. The parameter $r_{\rm p}$ must be kept small, so that the inverse Compton spectra are below the INTEGRAL upper limits. This is achieved at the expense of a relatively high Doppler factor, at $\doppler=65$. The brightness temperature at $32\,$GHz is $T_{\rm B}=2.5\times10^{12}\,$K. This model also shows the minimum value of $\numax$, found to be $\numax=1.5\times10^{15}\,$Hz.

The high-Doppler-factor model shows that if the emission at all frequencies originate from a single source region, it must be boosted by a much higher Doppler factor than proposed for the jet components by \citet{bachetal05}. Even with a viewing angle $\vartheta\approx0$, in which case $\doppler\approx2\Gamma$, a bulk Lorentz factor of $\Gamma>33$ is required, suggesting either that the source was travelling at a much higher speed than during the observations analysed by \citet{bachetal05} or that an interpretation of the superluminal motion in the VLBI jet that infers a much higher bulk Lorentz factor should be applied. One such suggestion is a conical jet \citep{gopalkrishnaetal06}. Alternatively, the jet components may decelerate as they travel down the jet \citep{marscher99,georganopouloskazanas03,ghisellinietal05}. The wide range of superluminal velocities shown in \citet{bachetal05} would then be a combined result of the variations in speed, as well as of the viewing angle.

\begin{figure}
\centering
\includegraphics[width=8cm]{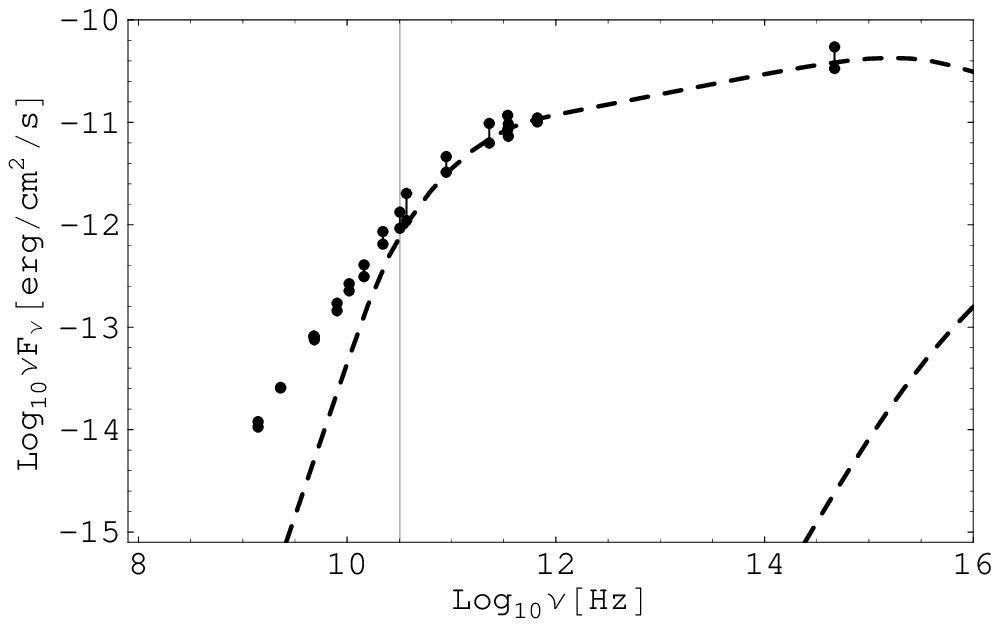}
\includegraphics[width=8cm]{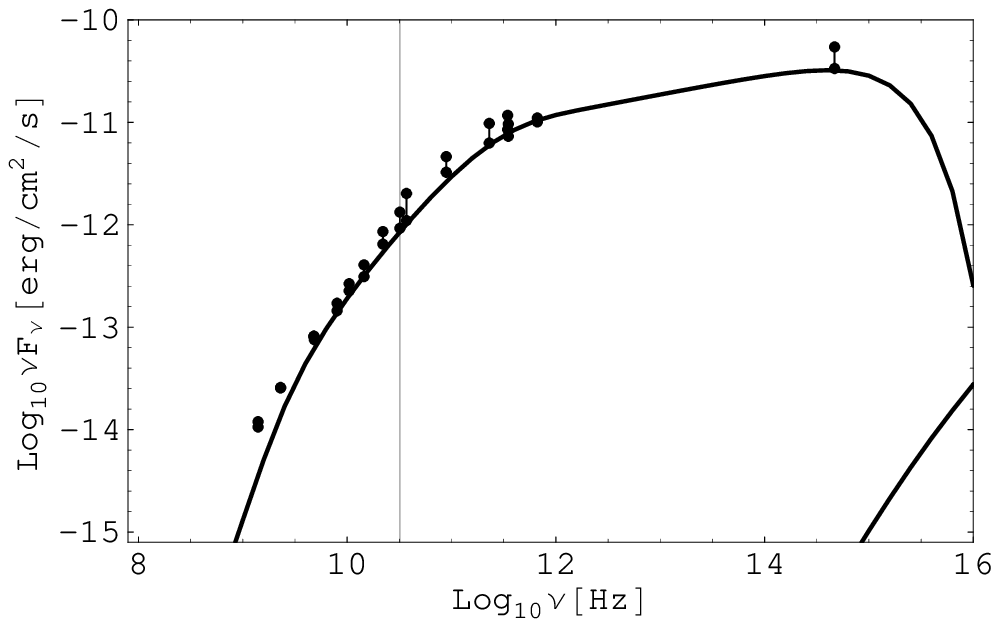}
\caption{The spectral energy distribution of S5~0716 +714 and the model spectra, as represented in Fig.~\ref{sedmono}, in the radio-to-optical band. Top panel shows the low-Doppler-factor model where the Doppler boosting factor is minimised. Bottom panel shows the high-Doppler-factor model where the values of the parameters are chosen to account for all radio and optical data points.}
\label{sedpowerlawopt}
\end{figure}

\section{Discussion}
\label{discussion}

\begin{figure}
\centering
\includegraphics[width=8cm]{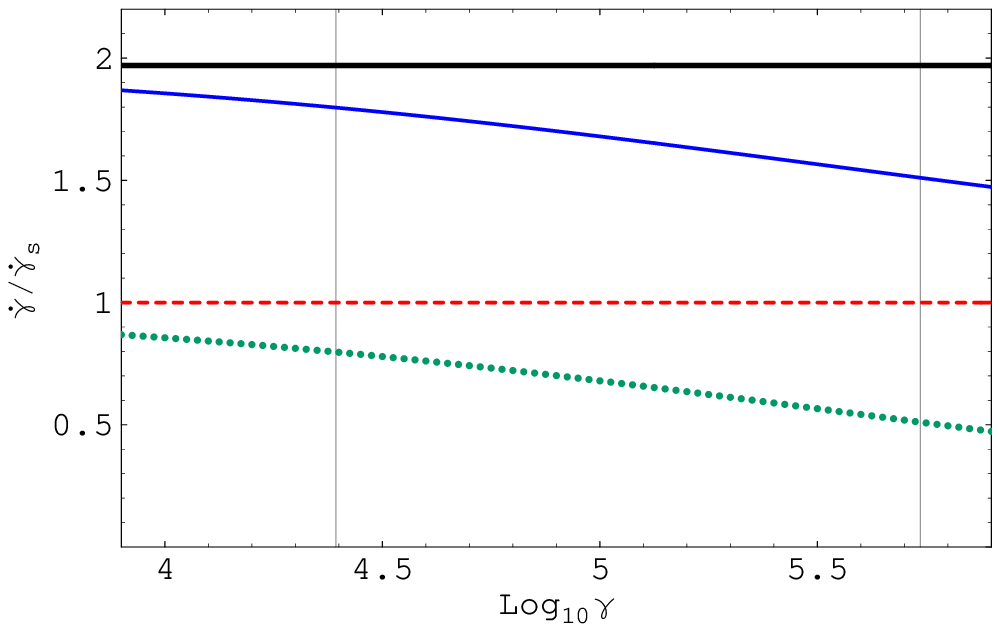}
\includegraphics[width=8cm]{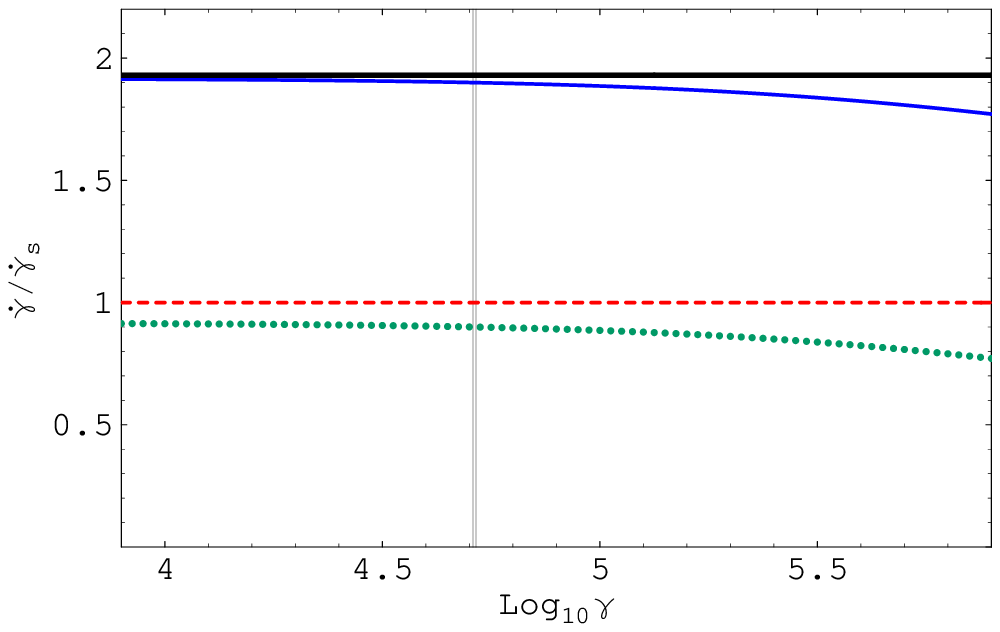}
\caption{Cooling rates normalised to the synchrotron loss rates (dashed lines at $\dot{\gamma}=\dot{\gamma}_{\rm s}$). The rate for inverse Compton scattering is shown as a dotted line, the total loss rate as a thin solid line (blue online). The approximate rate used to compute the electron distribution is shown as a thick solid line (black online).  The top (bottom) panel shows the results of the low(high)-Doppler-factor model shown as dashed (solid) lines in Fig.~\ref{sedpowerlawall}. The vertical dotted lines correspond to $\gamma=a\gcool$ (see Eq.~(\ref{elecspec})) on the left, and to $\gamma=\gammamax$ on the right.}
\label{figgammadot}
\end{figure}

The spectra of the four models --- two monoenergetic and two with double power-law electron injection --- shown in Figs.~\ref{sedmono}~and~\ref{sedpowerlawall}, show brightness temperatures at $32\,$GHz that are well above the conventional Compton limit. However, due to the lack of low-energy electrons, these brightness temperatures in fact lie below the threshold of the Compton catastrophe (i.e., $\xi<1$) and are, therefore, sustainable by the source.

An independent limit on $T_{\rm B}$ is provided by induced Compton
scattering, when low-energy electrons couple with high-frequency
photons. The photon occupation number $n_{\rm ph}(\nu)\propto
I_\nu/\nu^3$, which implies that the photon occupation
number is high at and below the peak (where $\nu=\nuabs$) of the synchrotron spectrum. In the presence of low-energy electrons, induced Compton scattering of photons at frequency $\nuabs$ to frequencies $\nu<\nuabs$ becomes increasingly significant as the synchrotron flux
grows. \citet{sunyaev71} showed that this process limits the
brightness temperature at a certain frequency to $T_{\rm B}<mc^2/(k_{\rm B}\tauT)=5\times10^9\,$K for
$\tauT\sim1$. \citet{sincellkrolik94} demonstrated by numerical
simulations that relativistic induced Compton scattering limits the
brightness temperature of a self-absorbed synchrotron source to
$T_{\rm B}<2\times10^{11}\nuGHz^{-1/(s+3)}\gamma_{\rm min}^{(s+2)/(s+5)}\,$K, where $\nuGHz$ is the observing frequency in units of GHz, $\gamma_{\rm min}$ is the low energy cut-off in the
electron spectrum $\propto \gamma^{-s}$. For a conventional power-law
electron spectrum in which $\gamma_{\rm min}=1$, this gives a limit of
$T_{\rm B}<2\times10^{11}\,$K at $1\,$GHz.

One might suspect that, at such high brightness temperatures as are
predicted by the models shown here, the effect of induced Compton
scattering should be significant. Qualitative arguments reveal the
contrary in our model, since the low-energy cut-off in the electron
spectrum is effectively $\gammap$, and the occupation number ($\propto
F_\nu/\nu^3$) of photons at frequencies that would couple with
electrons at $\gammap$ is negligibly small. Alternatively, when using the
result of \citet{sincellkrolik94}, one finds that the limit corresponds to
$T_{\rm B}<1.3\times10^{13}\nuGHz^{-2/11}(\gammap/10^3)^{3/5}\,$K for
$s=s_2\approx2.6$ and $\gammamin=\gammap$.

The spectral break at $\sim10^{11.5}\,$Hz was interpreted by \citet{ostoreroetal06} as the result
of the change in opacity of the source. In our interpretation, the break is a result of a corresponding spectral break in the electron spectrum; below this frequency, the synchrotron spectrum remains optically thin. The self-absorption frequency $\nuabs$ lies at a much lower frequency ($\sim4\,$GHz) in our model, which implies a weaker magnetic field and/or a lower electron density. This interpretation therefore does not require an inhomogeneous magnetic field and electron density.

The spectral index of the high-energy \lq\lq tail\rq\rq\ of the synchrotron
spectrum beyond the first spectral break depends on whether $\gammap$
lies below or above $\gcool$, since this affects the final shape of
the electron spectrum, as explained in Section~\ref{stationarysoln}. If the number of electrons leaving the
energy $\gammap mc^2$ is dominated by cooling by radiation, the synchrotron spectrum continues from $F_\nu\propto\nu^{1/3}$ between $\nuabs$ and $\nucool$, to $F_\nu\propto\nu^{-1/2}$ between $\nucool$ and $\nup$, then $F_\nu\propto\nu^{-s_2/2}$ between $\nup$ and
$\numax$, and it is cut off exponentially beyond $\numax$. In this case, the
low radio-frequency spectrum resembles that of a monoenergetic
electron distribution of energy $\gcool mc^2$. If, on the other hand,
losses are dominated by electrons evacuating the emission zone on a
time scale of $t_{\rm esc}=R/c$, the synchrotron spectrum then
continues from $F_\nu\propto\nu^{1/3}$ between $\nuabs$ and $\nup$, to
$F_\nu\propto\nu^{-(s_2-1)/2}$ between $\nup$ and $\nucool$, then
$F_\nu\propto\nu^{-s_2/2}$ between $\nucool$ and $\numax$, and it is again
cuts off exponentially beyond $\numax$. The low radio-frequency synchrotron spectrum can be approximated in the same way as the one from a monoenergetic distribution of electron of energy $\gammap mc^2$.

Observations of S5~0716+714 from infrared to optical frequencies
suggest that the spectral energy distribution between the frequencies
$\nu_{\rm K}=1.38\times10^{14}\,$Hz and $\nu_{\rm
  B}=6.81\times10^{14}\,$Hz can be well-fitted by the power law
$F_\nu\propto\nu^{-1.12}$ \citep{hagenthornetal06}. Clearly, the top
panel in Fig.~\ref{sedpowerlawopt} is much too hard at these
frequencies. However, the spectrum can be softened by lowering the
cut-off frequency of the synchrotron spectrum $\numax$ (i.e., bottom
panel in Fig.~\ref{sedpowerlawopt}). By decreasing $\numax$ to
approximately $\nu_{\rm K}$, the spectrum begins an exponential drop
at or just before reaching the relevant frequency range and, as a
result, softens the spectrum, without altering the level of flux or
the spectral shape at frequencies $<\numax$.

Figure~\ref{sedmono} and the bottom panel in Fig.~\ref{sedpowerlawopt}
demonstrate that, if the variability of S5~0716+714 is intrinsic and
extends down to $<32\,$GHz, the Doppler-boosting factor of the
emission region must be much higher than the $20-30$ suggested by
\citet{bachetal05}. One possibility for enabling the
Doppler factor to fall within this range is to increase the magnetic
field strength and the electron density, as shown in
Table~\ref{tabvalues}. This causes the synchrotron self-absorption
frequency $\nuabs$ to increase, as shown by the top panel in
Fig.~\ref{sedpowerlawopt}, therefore requiring the assumption that the
emission at frequency below $\nuabs$ originates from a larger region
than inferred from the variability at $32\,$GHz and
$37\,$GHz. Requiring the value of $\nuabs\leq32\,$GHz, we find that, to remain below the INTEGRAL upper limits, we require a minimum Doppler factor of $\doppler=30$ and a minimum self-absorption
frequency of $\nuabs=32\,$GHz. Below these minima, the model spectrum
would either have a level of flux below the measured flux at radio
frequencies or the subsequent first inverse Compton spectrum would
lie above the INTEGRAL upper limits.

The models shown here depart from the equipartition of magnetic and
particle energy, and are dominated by the energy of the relativistic
electrons (except for the rejected model). Estimating the energy
required by the source from the host galaxy, $P_{\rm jet}$, we find
that the power required from the source by our models is
consistent with what is expected from a low-energy peaked BL~Lac object
\citep[see e.g.,][]{nieppolaetal06}.

An approximation inherent in our method is that, as far as their effect on the electron distribution is concerned, inverse Compton losses are dominated by single scatterings off the synchrotron photons and may be treated in the Thomson approximation. For each of the models presented in Fig.~\ref{sedpowerlawopt}, we show in Fig.~\ref{figgammadot} the total rate of radiative cooling of an electron (thin solid lines, blue online) computed using the full Klein-Nishina loss rate for scattering on the full output spectrum. This should be compared with the thick solid lines (black online) that give the loss rates used in computing the electron density according to Eq.~(\ref{kineq}). These lines are horizontal and give the quantity $\dot{\gamma}_{\rm total}/\dot{\gamma}_{\rm s}= 1+\delta$ for the converged model solutions.  The range over which cooling is important in our approximate solution lies between the vertical lines at $\gamma=a\gcool$ and $\gammamax$ in this figure. (In the lower panel, these lines lie close together.) The maximum deviation in this range is roughly 20\% and occurs at $\gamma=\gammamax$ for the model with a relatively low Doppler factor (plotted as a dashed line in Fig.~\ref{sedpowerlawopt}). The deviation for the higher Doppler factor model is less than 3\%. We conclude that this approximation has a negligible effect on the electron distribution. The computation of the output spectrum from the electron distribution is not affected by this approximation, since
the full Klein-Nishina expression for the emissivity is used.

A testable prediction of our models is correlated variability. The
hard $\alpha\approx -0.3$ radio spectral component between $30\,$GHz
and several hundred GHz is interpreted as the optically thin emission
of electrons whose characteristic synchrotron frequency lies at or
above hundreds of GHz. Consequently, fluctuations in the number of
such particles should be simultaneously reflected in broad-band
fluctuations of the specific intensity of radiation at these
frequencies. We are not aware of studies that investigate such an
effect.

The variability of the synchrotron spectrum should also be correlated
to the variability of the inverse Compton spectrum, since the same
electrons are responsible for both emissions. Analysis of XMM
observation by \citet{ferreroetal06} show correlated variability
between two X-ray spectral components of S5~0716+714, 
which were interpreted as the high-energy part of the synchrotron spectrum and
the low-energy part of the inverse Compton spectrum. In the context of
our model, the synchrotron component is the emission from the high-energy tail of the electron spectrum, whereas the IC component is the synchrotron emission of the hard low-energy electron spectrum
scattered by the same low energy electrons. The study by
\cite{ferreroetal06} therefore does not provide a direct 
test of our model, although their data support an SSC interpretation. 

Similarly, the polarisation properties of the $30$ -- $100\,$GHz
emission are predicted to be those of the single-particle synchrotron
emission. One aspect of this is the possibility that the intrinsic
circular polarisation of the source might reach a few percent
\citep{kirktsang06}. Another concerns the degree of linear
polarisation. 
Power-law electron distributions radiating in a source
with a completely homogeneous magnetic field can theoretically reach
quite high degrees of linear polarisation (up to $\approx70\%$,
depending weakly on the power-law index). These values are not reached
in extra-galactic radio sources, probably because the magnetic field
direction within the source is tangled. In general, higher degrees of 
polarisation are found at higher frequency, consistent with the 
conventional picture in which the source size decreases with increasing 
frequency. 

In our model, however, the degree of 
linear polarisation in the radio is determined by monoenergetic electrons. 
According to standard synchrotron theory, this 
tends to $50\%$ at low frequency
($x=\nu/\nusynch\ll1$), rises to $76\%$ at $x=1$, and tends to $100\%$ 
at $x\gg1$. The effect of a tangled field within the source reduces
these values, but is independent of observing frequency. Thus, as in 
conventional inhomogeneous models, the degree of linear polarisation 
is predicted to increase
with frequency. However, more quantitative predictions would 
require consideration of effects, such as internal Faraday rotation,
and lie outside the scope of the present paper.

Finally, we note that the hard electron injection spectrum that we
have adopted ($\diffn\propto \gamma^2$) corresponds to the low-energy
($\gamma mc^2<k_{\rm B}T$) part of a relativistic thermal
distribution.  According to Table~1, the corresponding electron
temperature would lie at around $10^{12}\,$K. In an electron-ion
plasma, this corresponds to the temperature of shocked gas behind a 
mildly relativistic shock front, if one assumes that the electrons 
and ions equilibrate to a 
common temperature. Recent P.I.C. simulations suggest that 
this assumption may indeed be justified \citep{spitkovsky07}. 

\section{Conclusion}
\label{conclusion}

Using the specific case of S5~0716+714 as an example, we confirm that it
is possible to produce high brightness temperatures at GHz frequencies
in compact radio sources without the onset of catastrophic cooling,
provided that the radiating particles have a distribution
that is sufficiently hard below a characteristic. In addition, we show qualitatively
that induced Compton scattering is insignificant in sources with a low-energy electron cut-off despite the high brightness temperature, the underlying reason being the low occupation number of the photons that can couple with the electrons at the cut-off energy.

The model where an electron distribution that is a double power law
in energy, peaking at $\gammap$, is injected into the source offers
more flexibility at higher frequencies in the synchrotron spectrum
(from infrared to optical) at the expense of more free parameters,
compared to either monoenergetic or single power-law
distributions. These parameters should be constrained by simultaneous
observations due to the highly variable nature of IDV sources. In the
case of S5~0716+714 where such data is available, the spectral break
at about $230\,$GHz determines the value of $\gammap$, the optical
data at $5\times10^{14}\,$Hz gives the lower limit of $\numax$, as well
as constraining the spectral index $s_2$, and the INTEGRAL upper
limits give the upper limit of $\numax$ and also constrain the value
of $r_{\rm p}$, which in turn determines the electron density.

The example of S5~0716+714 illustrates several important spectral
properties of an electron distribution with a low-energy
cut-off, as described in the previous sections. The most noticeable
feature is the hard, inverted, optically thin synchrotron spectrum,
spanning a wide frequency range, which is a
prevalent feature in compact radio sources at radio frequencies
\citep[e.g.,][]{gearetal94,kedziorachudczeretal01}. Other features are
the spectral breaks at $\nup=\gammap^2\nu_0$,
$\nucool=\gcool^2\nu_0$, and the exponential cut-off at
$\numax=\gammamax^2\nu_0$. This model, therefore, allows a simple
homogeneous source to reproduce the common features shown by many IDV
sources.

\begin{acknowledgement}
We thank Luisa Ostorero and Stefan Wagner for helpful discussions and
for providing us with easy access to the observational data. We would also like to thank the anonymous referee for constructive comments and suggestions that we feel have led to a significant
improvement in this paper.
\end{acknowledgement}

\bibliographystyle{aa}
\bibliography{7962}

\end{document}